\pdfoutput=1

\documentclass[11pt]{article}

\usepackage[]{EMNLP2023}

\usepackage{times}
\usepackage{latexsym}
\usepackage{booktabs}
\usepackage{multirow}
\usepackage{makecell}
\usepackage[T1]{fontenc}
\usepackage{xcolor,soul}
\usepackage{arydshln}
\usepackage [normalem]{ulem}

\usepackage[utf8]{inputenc}

\usepackage{microtype}

\usepackage{inconsolata}

\usepackage{tabu}                      
\usepackage{booktabs}                  
\usepackage{lipsum}                    
\usepackage{mwe}                       
\usepackage{algorithm}
\usepackage{algpseudocode} 
\usepackage{amsmath} 
\usepackage{amsfonts}

\usepackage{mathptmx}                  

\usepackage{soul}
\usepackage{makecell}
\title{LLM4Vis: Explainable Visualization Recommendation using ChatGPT}

\author{
~~Lei Wang$^\spadesuit$
~~Songheng Zhang$^\spadesuit$
~~Yun Wang$^\diamondsuit$
~~Ee-Peng Lim$^\spadesuit$
~~Yong Wang$^\spadesuit$\thanks{~~Corresponding author.} \\
$^\spadesuit$Singapore Management University\\ 
$^\diamondsuit$Microsoft Research Asia\\
\small{\tt \{lei.wang.2019, shzhang.2021, eplim, yongwang\}@smu.edu.sg} \quad {\tt wangyun@microsoft.com} \\
}

\begin{document}
\maketitle
\begin{abstract}
Data visualization is a powerful tool for exploring and communicating insights in various domains. To automate visualization choice for datasets, a task known as visualization recommendation has been proposed. Various machine-learning-based approaches have been developed for this purpose, but they often require a large corpus of dataset-visualization pairs for training and lack natural explanations for their results.
To address this research gap, we propose LLM4Vis, a novel ChatGPT-based prompting approach to perform visualization recommendation and return human-like explanations using very few demonstration examples. 
Our approach involves feature description, demonstration example selection, explanation generation, demonstration example construction, and inference steps.
To obtain demonstration examples with high-quality explanations, we propose a new explanation generation bootstrapping to iteratively refine generated explanations by considering the previous generation and template-based hint.
Evaluations on the VizML dataset show that LLM4Vis outperforms or performs similarly to supervised learning models like Random Forest, Decision Tree, and MLP in both few-shot and zero-shot settings. The qualitative evaluation also shows the effectiveness of explanations generated by LLM4Vis.
We make our code publicly available at \href{https://github.com/demoleiwang/LLM4Vis}{https://github.com/demoleiwang/LLM4Vis}. 
\end{abstract}

\section{Introduction}

Data visualization is a powerful tool for exploring data, communicating insights, and making informed decisions across various domains, such as business, scientific research, social media and journalism~\cite{munzner2014visualization,ward2010interactive}. However, creating effective visualizations requires familiarity with data and visualization tools, which can take much time and effort~\cite{data2vis2019dibia}.  A task that automates the choice of visualization for an input dataset, also known as \textit{visualization recommendation}, has been proposed.  

So far, visualization recommendation works can be categorized into rule-based and machine learning-based approaches~\cite{vizml2019hu, li2021kg4vis, zhang2023adavis}. 
Rule-based approach~\cite{apt1986machinlay, vartak2015seedb, demiralp2017foresight} leverages data characteristics and visualization principles to predict visualizations, but suffers from the limited expressibility and generalizability of rules. Machine learning-based approach~\cite{vizml2019hu, wongsuphasawat2015voyager, zhou2021table2charts} learns machine learning (ML) or deep learning (DL) models from dataset-visualization pairs and these models can offer greater recommendation accuracy and scalability.
Existing ML/DL models, however, often need a large corpus of dataset-visualization pairs in their training and they could not provide explanations for the recommendation results.
Recently, a machine learning-based work, KG4Vis~\cite{li2021kg4vis}, leverages knowledge graphs to achieve explainable visualization recommendation. Nevertheless, KG4Vis still requires supervised learning using a large data corpus and its explanations are generated based on predefined templates, which constrain the naturalness and flexibility of explanations.

Recently, large language models (LLMs) such as ChatGPT~\cite{openai-chatgpt-2022} and GPT-4~\cite{openai-gpt4-2023} have demonstrated strong reasoning abilities using in-context learning~\cite{gpt3, opt, chowdhery2022palm}.
The key idea behind this is to use analogical exemplars for learning~\cite{dong2022survey}. 
Through in-context learning, LLMs can effectively perform complex tasks, including but not limited to mathematical reasoning~\cite{cot_wei}, visual question answering~\cite{yang2022empirical_icl_mm}, and tabular classification~\cite{hegselmann2023tabllm} without supervised learning.
By prompting the pretrained LLM to perform tasks using in-context learning, we avoid the overheads of parameter updates when adapting the LLM to a new task.

Inspired by the excellent performance of ChatGPT on natural language tasks~\cite{Qin2023-ar,Li_undated-wy,Sun2023-be,Gilardi2023-qo}, we explore the possibility of leveraging ChatGPT for explainable visualization recommendation.
Specifically, we propose \textit{LLM4Vis}, a novel ChatGPT-based In-context Learning approach for
\underline{Vis}ualization recommendation with natural human-like explanations by learning from very few dataset-visualization pairs.
LLM4Vis consists of several key steps: feature description, demonstration example selection, 
explanation generation bootstrapping, prompt construtction,  
and inference for explainable visualization recommendation.
Firstly, feature description is used to quantitatively represent the characteristics of tabular datasets, which makes it easier to analyze and comprehend tabular datasets using ChatGPT.
Demonstration example selection is then employed to prevent the input length from exceeding the maximum length of ChatGPT by retrieving $K$ nearest labeled data examples. 
Next, we propose a new iterative refinement strategy in terms of the previous generation and hint to obtain a more high-quality recommendation explanation and a score of each visualization type before prompt construction.
Finally, the constructed prompt is used to guide ChatGPT to recommend visualization types for a test tabular dataset while providing recommendation scores and human-like explanations. 
We evaluate the visualization recommendations of LLM4Vis by comparing its accuracy of visualization with strong machine learning-based baselines from VisML~\cite{hu2019vizml} like Decision Trees, Random Forests, and MLP. 
The visualization recommendation results demonstrate that LLM4Vis outperforms all the baselines in few-shot and full-sample training settings.
Furthermore, 
the evaluations conducted by LLM and humans show that the generated explanation of the test data example matches the predicted score.
Our contributions are summarized below:
\begin{itemize}
    \item We present LLM4Vis, a novel ChatGPT-based prompting approach for visualization recommendation, which can achieve accurate visualization recommendations with human-like explanations.

    \item We propose a new explanation generation bootstrapping method to generate high-quality recommendation explanations and scores for prompt construction.
    
    \item 
    Experiment results show the usefulness and effectiveness of LLM4Vis, encouraging further exploration of LLMs for visualization recommendations. 
\end{itemize}


\begin{figure*}[t]
    \centering
    \includegraphics[width=1.0\linewidth]{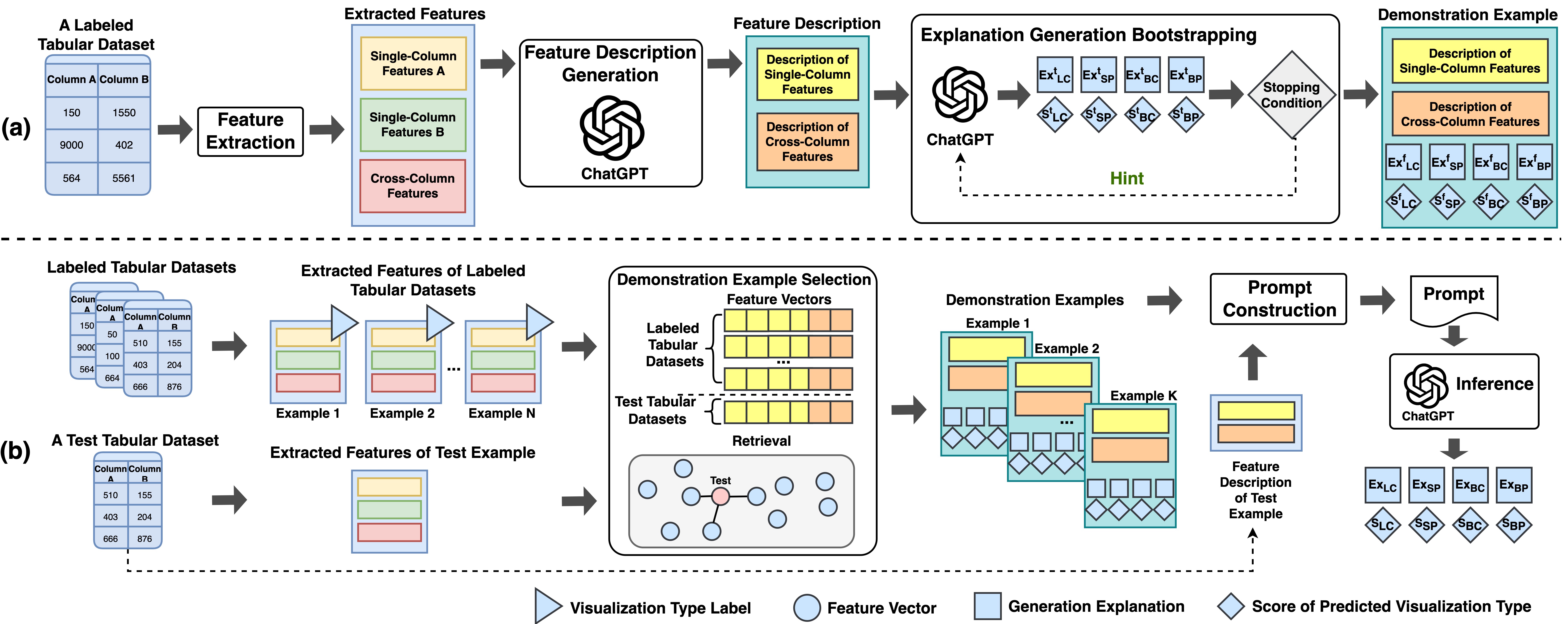}
    \vspace{-20pt}
    \caption{A detailed illustration of LLM4Vis. (a) The process for converting a labeled tabular dataset to a demonstration example of the final prompt, including feature extraction, feature description, and explanation generation bootstrapping. (b) The process for visualization type recommendation of a test tabular dataset, involving demonstration example selection, prompt construction, and inference. 
    }
    \label{fig:overview}
    \vspace{-15pt}
\end{figure*}
\section{Related Work}

Prior studies on automatic visualization recommendation approaches can be categorized into two groups: unexplainable visualization recommendation approaches and explainable visualization approaches~\cite{wang2021survey}.
Unexplainable visualization recommendation approaches,
including Data2vis~\cite{dibia2019data2vis}, VizML~\cite{hu2019vizml}, and Table2Chart~\cite{zhou2021table2charts}, 
can recommend suitable visualizations for an input dataset, but cannot provide the reasoning behind the recommendation to users, making them black box methods.
Explainable visualization recommendation approaches provide explanations for their recommendation results, enhancing transparency and user confidence in the recommendations.
Most rely on human-defined rules, such as Show Me~\cite{mackinlay2007show} and Voyager~\cite{wongsuphasawat2015voyager}.
But rule-based approaches are often time-consuming and resource-intensive, and require visualization experts' manual specifications. To address such limitations, \citet{li2021kg4vis} proposed a knowledge graph-based recommendation method (KG4Vis) that learns the rules from existing visualization instances. 
To provide human-like explanations, this paper proposes to leverage ChatGPT to recommend appropriate visualizations.
\section{LLM4Vis Method}

\subsection{Overview}
In this section, we present the proposed approach LLM4Vis. 
As shown in Figure~\ref{fig:overview}, LLM4Vis consists of several key steps: feature description, demonstration example selection, explanation generation bootstrapping, prompt construction, and inference. 
To save space, we show the exact wording of all prompts we employ in LLM4Vis in the Appendix.

\subsection{Feature Description}
Most large language models, such as ChatGPT~\cite{openai-chatgpt-2022}, are trained based on text corpora. To allow ChatGPT to take a tabular dataset as input, we can first use predefined rules to transform it into sets of data features that quantitatively represent its characteristics. Subsequently, these features can be serialized into a text description.

Following VizML~\cite{vizml2019hu} and KG4Vis~\cite{li2021kg4vis}, we extract 80 \textit{cross-column} data features that capture the relationships between columns and 120 \textit{single-column} data features that quantify the properties of each column. 
We categorize the data features related to columns into \textit{Types}, \textit{Values}, and \textit{Names}. Types correspond to the columns' data types, Values capture statistical features such as distribution and outliers, and Names are related to columns' names.

Previous works~\cite{hegselmann2023tabllm, dinh2022lift} perform serialization mainly through the use of rules, templates, or language models. 
In this paper, to ensure grammatical correctness, flexibility, and richness, we follow the LLM serialization method proposed by TabLLM~\cite{hegselmann2023tabllm}.
Specifically, our approach involves providing a prompt that instructs ChatGPT to generate for each tabular dataset a comprehensive text description that analyzes the feature values from both single-column and cross-column perspectives. 
The feature description is then used to construct  concise but informative demonstration examples.

\subsection{Demonstration Example Selection}

Due to the maximum input length restriction, a ChatGPT prompt could only accommodate a small number of demonstration examples. The selection of good demonstration samples from a large set of labeled data is therefore crucial. Instead of randomly selecting examples that may not be relevant to the target test tabular dataset~\cite{liu2021makes},
we first represent each tabular dataset by converting its features to a vector. 
Then, we use a clustering algorithm to select a representative subset of examples from the labeled set. The clustering algorithm creates $C$ clusters, and we choose $R$ representative examples from each cluster, resulting in a subset of size $M = C \times R$ as the retrieval set.
Finally, we retrieve $K$ training data examples with the highest similarity scores with a target data example based on the cosine similarity scores of their vector representations from the retrieval set. 

\subsection{Explanation Generation Bootstrapping}
Each labeled data example $X_i$ comes with only one ground truth label $Y_i$, but not the explanation required to be used in a demonstration example. 
We therefore propose a prompt to leverage the built-in knowledge of ChatGPT to recommend the appropriate visualization and the corresponding explanation
for each labeled dataset. 
Our strategy involves instructing ChatGPT to generate a response in a JSON format, where the keys correspond to four possible visualization types $\{Y_{LC}, Y_{SP}, Y_{BC}, Y_{BP}\}$ ($LC$: line chart, $SP$: scatterplot, $BC$: bar chart, $BP$: Box plot) and the values are recommendation scores $\{S_{LC}, S_{SP}, S_{BC}, S_{BP}\}$.
Furthermore, we prompt ChatGPT to generate explanations $\{Ex_{LC}, Ex_{SP}, Ex_{BC}, Ex_{BP}\}$ for its prediction of each visualization type in an iterative process.

Specifically,
we employ zero-shot prompting with the feature description of a tabular dataset to ask ChatGPT to generate scores $\{S^1_{LC}, S^1_{SP}, S^1_{BC}, S^1_{BP}\}$ for all visualization types and provide explanations $\{Ex^1_{LC}, Ex^1_{SP}, Ex^1_{BC}, Ex^1_{BP}\}$ supporting these scores' assignment to each visualization type. The sum of these scores is required to be 1.
Subsequently, these scores and explanations are revised by an iterative refinement process that terminates when the ground truth visualization type $Y_i$ receives the highest score which also exceeds the second-highest score by at least a margin of $0.1$. 
The final explanations and scores are denoted by $\{Ex^f_{LC}, Ex^f_{SP}, Ex^f_{BC}, Ex^f_{BP}\}$ and scores $\{S^f_{LC}, S^f_{SP}, S^f_{BC}, S^f_{BP}\}$.
However, if the ground truth visualization type does not meet the aforementioned conditions, we develop a hint and append it to the initial zero-shot prompting to instruct ChatGPT to produce a more accurate output. An example hint template is as follows: ``\textit{\{a\} may be more suitable than \{b\}. However, the previous scores were \{c\}}''. The \textit{\{a\}} slot is for the ground truth label, the \textit{\{b\}} slot is for the incorrect label with the highest score, and the \textit{\{c\}} slot is for the previously predicted score for each visualization type.
In the Experiment section, we compare two hint strategies, including using ground truth (GT-As) and random labels (Rand-As) as hints. The results can be found in Figure~\ref{fig:ablation}.

Through this iterative refinement, we can obtain higher-quality visualization type prediction with scores and corresponding explanations. 
Note that if the labeled dataset fails to meet the stopping condition within the maximum iteration steps, we will delete this data example from the retrieval set.

\subsection{Prompt Construction and Inference}

After retrieving $K$ nearest labeled samples 
from the retrieval set for a test data sample, along with their
feature descriptions, refined explanations, and refined scores, 
each demonstration example is constructed with the feature description, task instruction, recommended visualization types with scores, and explanations.
Then, we incorporate the feature description of a test data example into a pre-defined template. 
Next, the constructed demonstration examples and the completed template for the test data example are concatenated and fed into ChatGPT to perform visualization type recommendations.
Finally, we extract the recommended visualizations and explanations from the ChatGPT output.

\section{Evaluation}

\subsection{Evaluation Setup}

\begin{table}[t]
    \centering
    \caption{
    The result of our quantitative evaluation with the best results highlighted in bold.
    LLM4Vis-random refers to randomly selecting demonstration examples from the retrieval set. 
    Conversely, LLM4Vis-retrieval refers to retrieve $K$ nearest labeled data examples from the retrieval set.
    Note that LLM4Vis using 5 demonstrations shows a performance better than machine learning based baselines trained with full samples (5000) and provides human-like explanations that are unattainable with these baselines. 
    }
    \vspace{-5pt}
    \resizebox{0.48\textwidth}{!}{
     \begin{tabular}{ll|ccccc}
    \toprule  
     \bf Settings &\bf Methods &  \multicolumn{5}{c}{\bf Hits@2}  \\
      & & \bf Line& \bf Scatter&\bf Bar&\bf Box&\bf Overall \\
     
     \midrule
    \bf \multirow{3}{*}{Full Samples}& Decision Tree& 57.3 & 60.0  & \bf100 & 56.0 & 68.3  \\
    &Random Forest & 92.0 & \bf100  & 90.7 & 32.0 & 78.7   \\
    &MLP &\bf 97.3 & \bf100  & 93.3 & 24.0 & 78.7 \\
    \midrule
    \bf \multirow{4}{*}{\makecell{Few-Shot (4) \\ Fixed}}& Decision Tree& 42.7 & 12.0  & \bf100 & 41.3 & 49.0   \\
    &Random Forest & 66.7 & 78.7  & 38.7 & 65.3 & 62.0   \\
    &MLP & 70.7 & 85.3  & 44.0 & 45.3 & 61.0   \\
    &\textbf{LLM4Vis} & 53.3 & 80.0  & 84.0 & 93.3 & 77.7   \\
    \midrule
    \bf \multirow{4}{*}{\makecell{Few-Shot \\ Dynamic}} &LLM-SP-Random & 36.0 & 86.0  & 96.0 & 46.0 & 66.0  \\
    &LLM-SP-Retrieval & 68.0 & 94.0  & 90.0 & 32.0 & 71.0  \\
    &{LLM4Vis-Random} & 46.7 & 69.3  & 84.0 & 90.7 & 72.7  \\
    &\textbf{LLM4Vis-Retrieval} & 62.4 & 96.0 & 86.8 & \bf97.2 & \bf85.7  \\
    \midrule
    
    \bf \multirow{2}{*}{Zero-Shot}& LLM-SP& 64.0 & 84.0  & 56.0 & 64.0 & 65.0  \\
    &\textbf{LLM4Vis}& 64.0 & 88.0  & 76.0 & 89.3 & 79.3  \\
     
    \bottomrule
    \end{tabular}
  
    }
    
    \label{tab:main_result}
    \vspace{-10pt}
\end{table}


\textbf{Dataset.}
We utilize the VizML corpus~\cite{vizml2019hu} to construct our training, validation, and test sets. 
We select a subset of 100 data-visualization pairs from the corpus to evaluate our model's performance for testing purposes. 
These pairs comprised 25 line charts, 25 scatter plots, 25 bar charts, and 25 box plots. 
We employ two different training settings for our experiments. In the first setting, we use the set of 5000 data-visualization pairs from the corpus to train all baseline models. In the second few-shot setting, we employ clustering techniques~\cite{Pedregosa2011ScikitlearnML} to extract $4\times 15$ data-visualization pairs from the 5000 pairs to build the retrieval set of size ($M=60$).

\noindent\textbf{Large Language Model Setup.}
We conduct experiments using the gpt-3.5-turbo-16k version of GPT-3.5, widely known as ChatGPT. We have chosen ChatGPT because it is a publicly available model commonly used to evaluate the performance of large language models in downstream tasks~\cite{Sun2023-be, Qin2023-ar, Li_undated-wy}. To conduct our experiments, we utilize the OpenAI API, which provides access to ChatGPT. Our experiments were done between June 2023 and July 2022, and the maximum number of tokens allowed for generation is set to be 1024. To enhance the determinism of our generated output, we set the temperature to 0. Due to the input length restriction of ChatGPT (i.e., 16,384 tokens), we limit the number of our in-context demonstrations $K$ to 8.

\noindent\textbf{Baselines.}
We compare
with strong visualization type recommendation baselines from VizML~\cite{hu2019vizml}. 
Specifically, we compare our method with Decision Tree, Random Forest, and MLP baselines, which are implemented using scikit-learn with default settings~\cite{Pedregosa2011ScikitlearnML}.
With full data training, these strong baselines are expected to outperform few-shot methods.
We also compare our method to a simple prompting technique named LLM-SP. In the zero-shot setting, the instruction in the prompting is to ask ChatGPT to recommend visualization type based on extracted features of the given tabular dataset. In the few-shot setting, each demonstration example in the prompt is composed of an instruction, extracted features of a given tabular dataset, and the corresponding labeled visualization type.


\noindent\textbf{Metrics.}
Our proposed method 
makes two visualization design choices based on the large language models directly. Referring to KG4Vis~\cite{li2021kg4vis}, we employ a commonly used metric to assess the effectiveness of our approach: \textit{Hits@2}, which indicates the proportion of correct visualization design choices among the top two options.

\subsection{Main Results}
Table~\ref{tab:main_result} shows that our few-shot LLM4Vis outperforms all baselines, including Decision Tree, Random Forest, and MLP, in the full sample training setting, which indicates that LLMs can effectively recommend appropriate visualization types by learning from limited demonstration examples and capitalizing on built-in background knowledge of visualization.
Note that even zero-shot LLM4Vis can outperform these strong baselines.
Two categories for few-shot settings are: \textit{fixed} and \textit{dynamic}.
In the fixed setting, fixed demonstration examples are chosen for all test examples, LLM4Vis outperforms all baselines.
In the dynamic setting, we select relevant demonstration examples for each test example.
LLM4Vis with dynamic few-shot settings
outperforms randomly selected demonstrations. It indicates that relevant demonstration examples can provide useful information to guide the LLM in recommending a suitable visualization type for the test tabular dataset.


\subsection{In-depth Analysis}

\begin{figure}[!htp]
    \centering
    \includegraphics[width=0.98\linewidth]{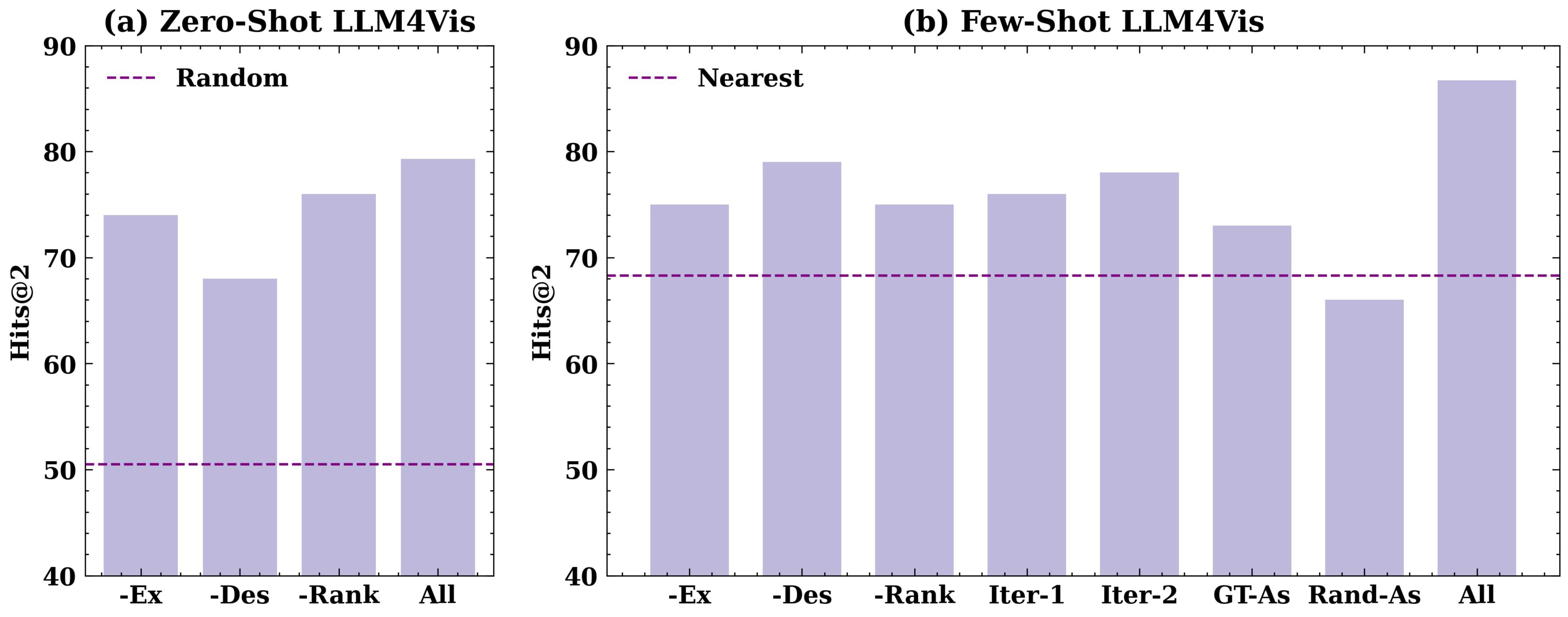}
    \vspace{-5pt}
    \caption{Effect of each component of LLM4Vis. All methods are evaluated on the same test dataset. 
    \textbf{All}: keeping all module unchanged. 
    \textbf{Random}: randomly choosing one visualization type as recommendation.
    \textbf{-Ex}: removing explanation in the prompt.
    \textbf{-Des}: removing feature description in the prompt.
    \textbf{-Rank}: predicting visualization type directly.
    \textbf{Nearest}: predicting using the nearest example.
    \textbf{Iter-1}: using explanation without refinement in the prompt. 
    \textbf{Iter-2}: using explanation with one step refinement in the prompt. 
    \textbf{GT-As}: generating the explanation in the prompt using the ground truth label as the hint.
    \textbf{Rand-As}: generating the explanation in the prompt using the random label as the hint.
    }
    \label{fig:ablation}
    \vspace{-10pt}
\end{figure}

\begin{figure}[t] 
    \centering
    \begin{tabular}{cc}
    \vspace{-3pt}
    \begin{tiny}\bf (a)\end{tiny} & \begin{tiny}\bf (b)\end{tiny}\\
	\includegraphics[width=0.44\linewidth]{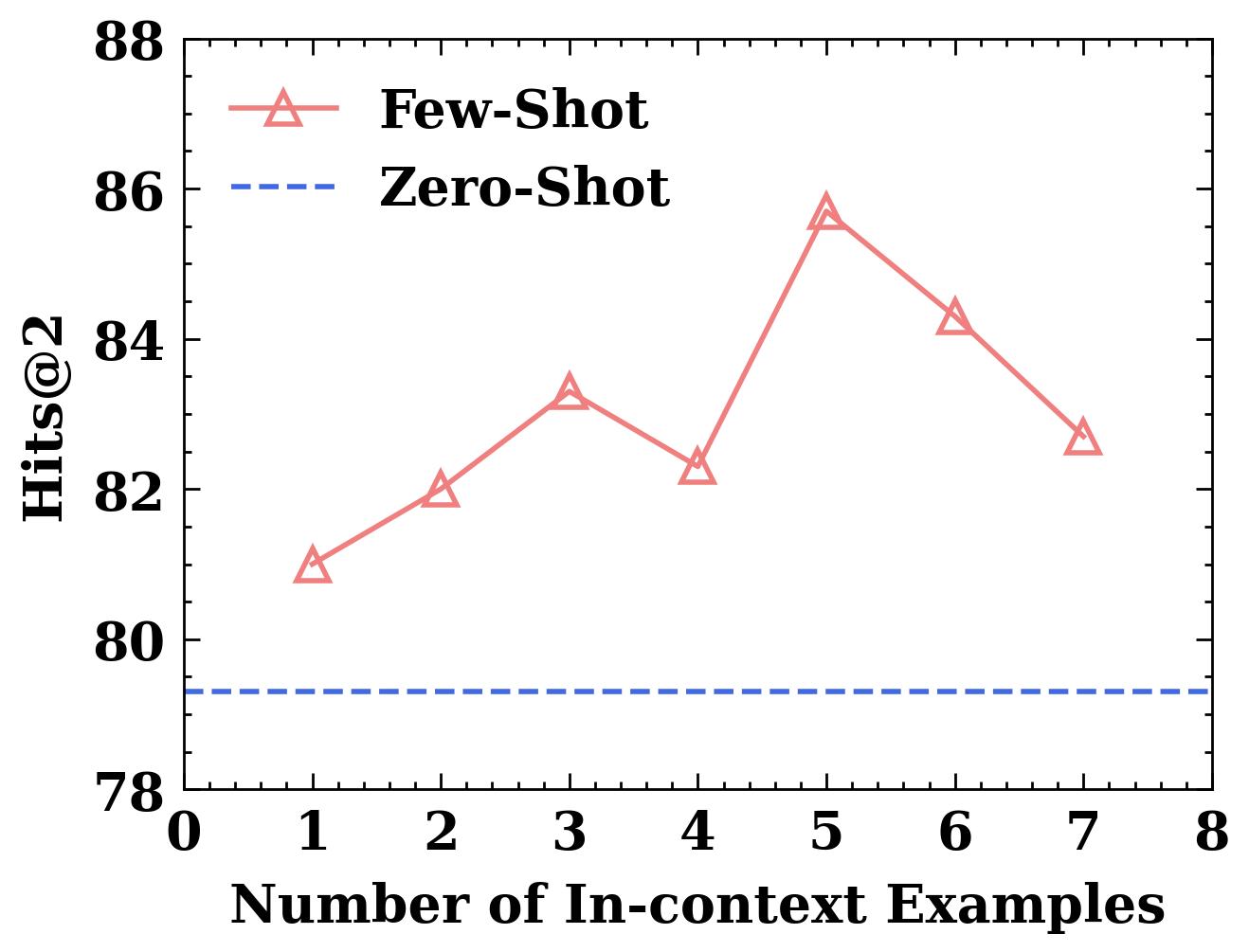}&
        \includegraphics[width=0.44\linewidth]{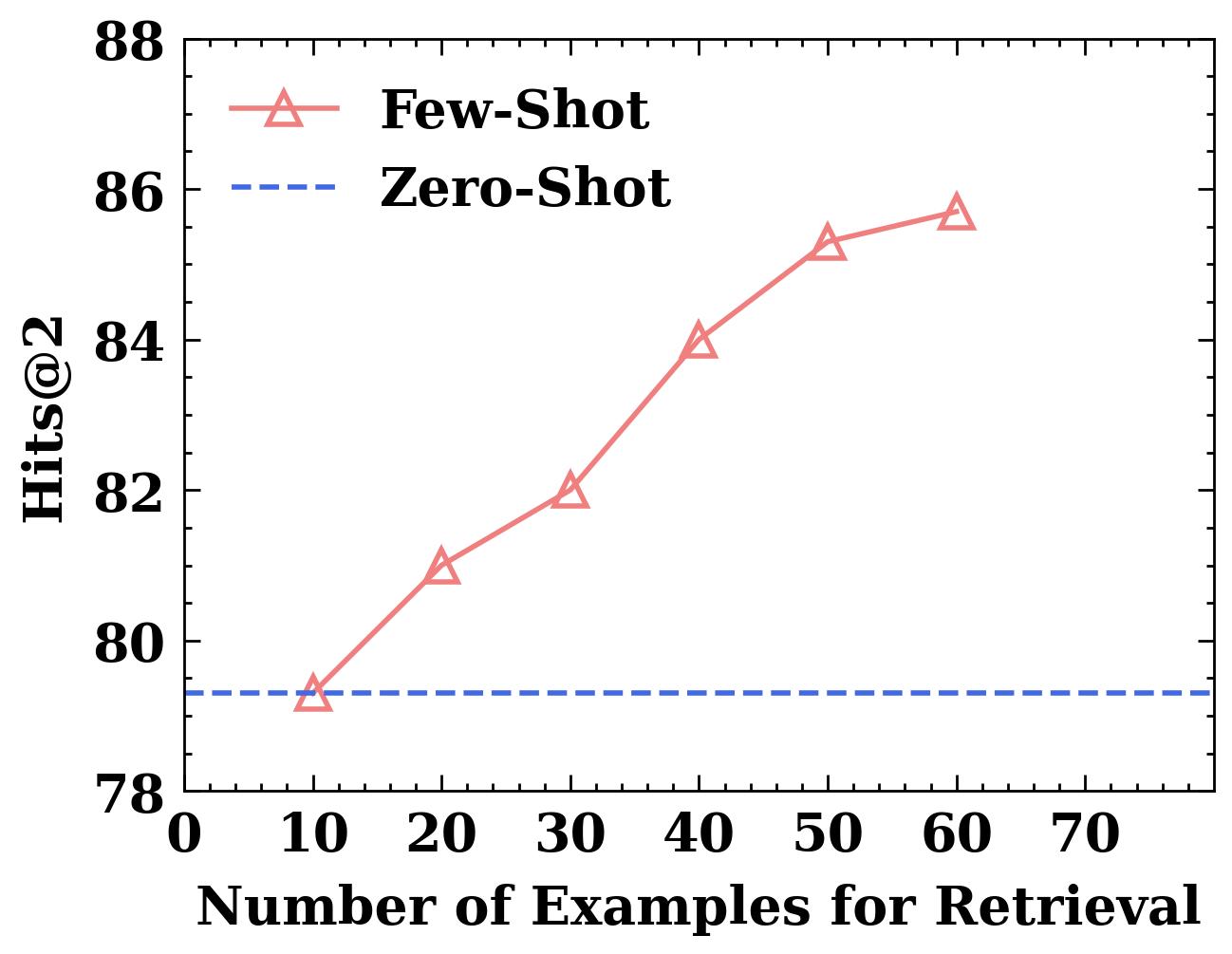} \\
        \vspace{-3pt}
        \begin{tiny}\bf (c)\end{tiny} & \begin{tiny}\bf (d)\end{tiny}\\
        \includegraphics[width=0.44\linewidth]{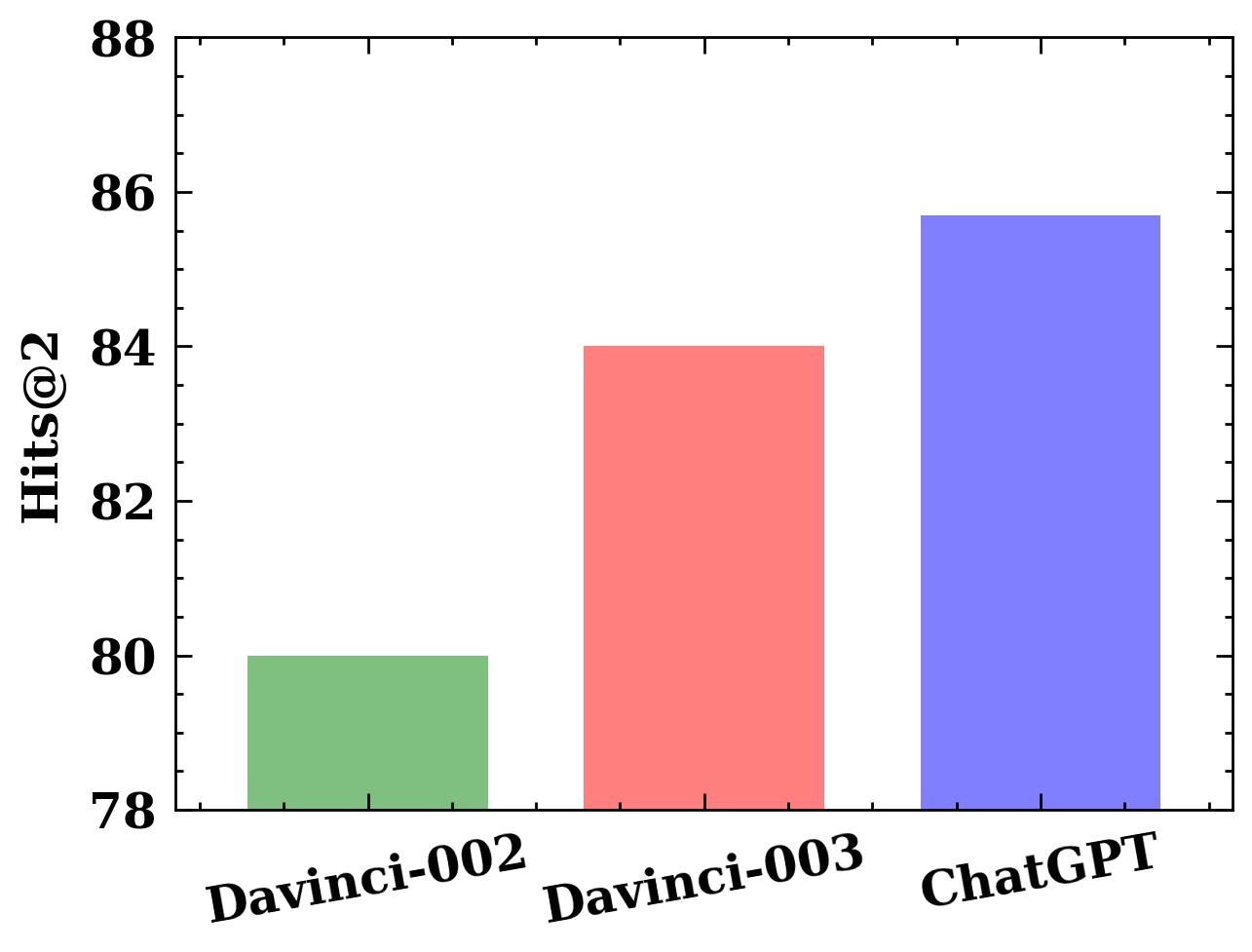} &
        \includegraphics[width=0.44\linewidth]{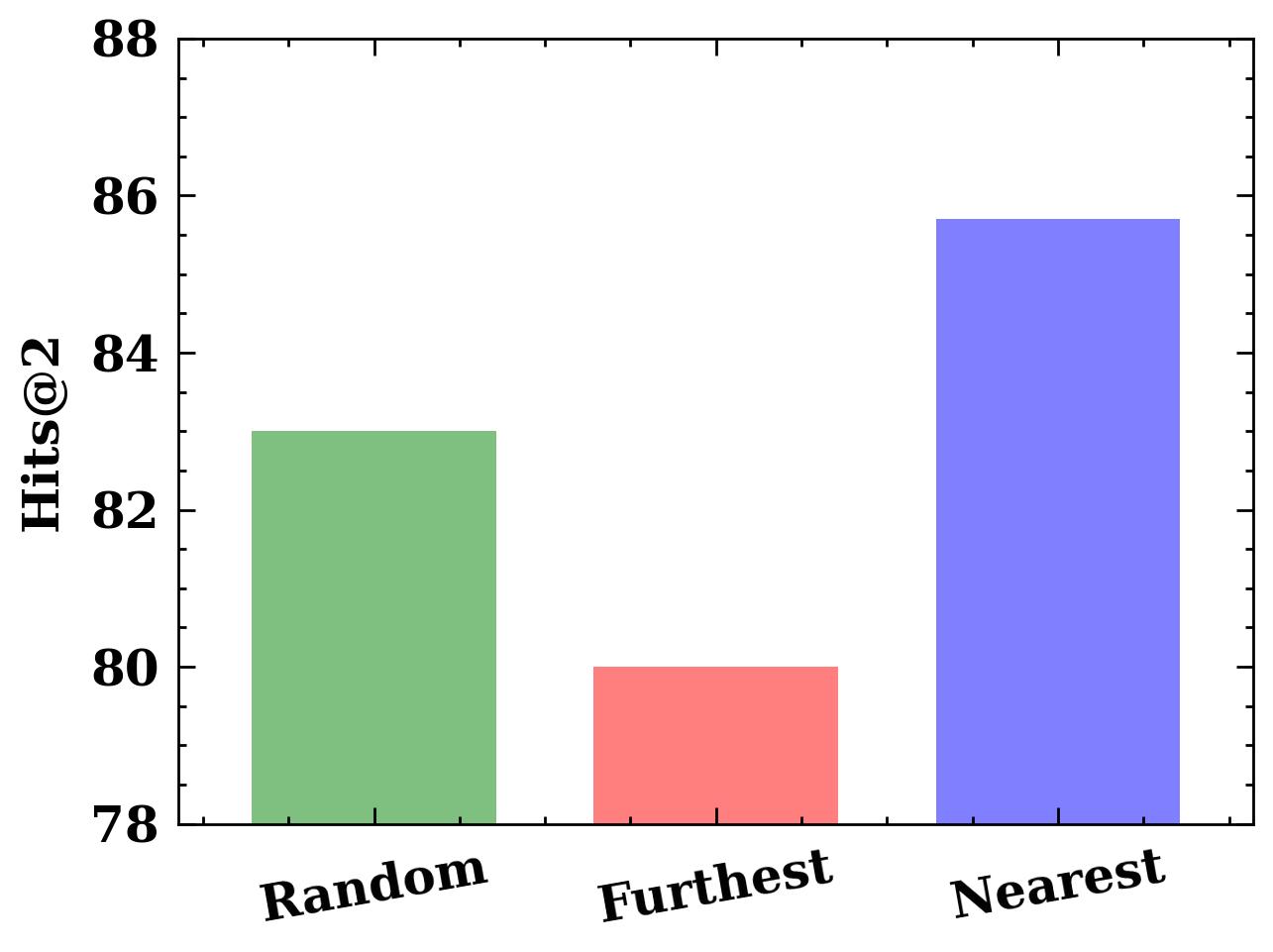} \\
    \end{tabular}
    \vspace{-5pt}
    \caption{Effect of the number of in-context examples (a), the number of examples in the retrieval set (b), different base large language model (c), and the ordering of K nearest examples as in-context examples (d).}
    \label{fig:further_analysis}
    \vspace{-10pt}
\end{figure}


\paragraph{Effect of each Component of LLM4Vis.}
Figure~\ref{fig:ablation} presents the comparison results of variants of LLM4Vis, wherein one component is either removed or replaced. The findings reveal that the absence of explanations, feature descriptions, and recommendation scores in the prompt consistently leads to reduced performance in both zero-shot and few-shot settings. With more iterations of explanation refinement, the performance improves. Replacing the proposed hint with the ground truth label or a random label results in a substantial drop in performance. Similarly, using the prediction from the nearest demonstration example as the test example's prediction also leads to significant performance degradation, which indicates that LLM effectively learns from given demonstration examples rather than merely copying them. Overall, all components of the proposed LLM4Vis contribute to recommendation accuracy.

\paragraph{Effect of the Number of In-context Examples.}
We assess the effect of the number of demonstration examples on LLM4Vis's performance. Specifically, we examine LLM4Vis, using different sets of nearest demonstration examples, ranging from 1 to 7 instances. The results, depicted in Figure~\ref{fig:further_analysis}(a), show that more demonstration examples lead to better performance, despite a drop when the number of demonstration examples goes from 3 to 4.

\paragraph{Effect of the Size of Retrieval Set.}
We quantify the impact of the size of the retrieval set. We test LLM4Vis on retrieval sets of varying sizes, ranging from $10$ to $60$ examples.
Figure~\ref{fig:further_analysis}(b) shows that the performance of LLM4Vis improves as the size of the retrieval set increases. This is likely because the larger retrieval set can find more relevant nearest neighbors. It indicates that LLM4Vis can achieve better results by scaling the retrieval set. 
As the retrieval set size increases from 50 to 60, we observe a decline in the degree of performance improvement. It suggests that the relevant information to test data in the k-nearest demonstration example may not have a proportional increase.

\paragraph{Effect of Base Large Language Models}
We also evaluate LLM4Vis using various LLMs, including different versions of GPT-3.5. According to official guidelines, ChatGPT has the highest capability, and \texttt{text-davinci-002} is the least capability model among the three LLMs. 
As expected, Figure~\ref{fig:further_analysis}(c) illustrates that model performance improves as the model capability increases from \texttt{text-davinci-002} to ChatGPT. 
Overall, these results indicate that LLMs of stronger capabilities usually deliver much better recommendation accuracy.

\paragraph{Effect of In-context Example Order.}
We compare three demonstration orders: random (shuffle $K$ nearest neighbors), 
furthest (samples with the least similarity are first selected), and nearest (samples with the most similarity are first selected).
The results in Figure~\ref{fig:further_analysis}(d) show that LLM4Vis is sensitive to the order of $K$ selected demonstrations. 
Specifically, employing the ``furthest'' ordering within the framework of LLM4Vis yields the lowest results, whereas the ``nearest" ordering yields the strongest performance. It indicates that relevant demonstrations can stabilize in-context learning of LLMs.

\paragraph{Explanation Evaluation.}
In this section, we assess the consistency between generated explanations and predicted scores of visualization type recommendations in a test tabular dataset. Two evaluation metrics are employed: LLM-based evaluation and human evaluation.

The LLM-based evaluation measures the Pearson correlation between the predicted scores generated by LLM4Vis and scores predicted by ChatGPT based on the explanations generated by LLM4Vis. A higher Pearson correlation signifies stronger consistency between the predicted scores and explanations. We obtain a Pearson correlation of 0.78 for zero-shot LLM4Vis and 0.92 for few-shot LLM4Vis. These findings indicate that the few-shot LLM4Vis exhibits greater consistency between its predicted scores and generated explanations than the zero-shot LLM4Vis.

Besides the LLM-based evaluation, we manually inspect ten correct recommendations to validate the consistency of generated explanations further and predicted scores. Our examination shows that nine out of the ten examples demonstrate consistent alignment between their explanations and predicted scores. 
The generated explanation and predicted score of one particular instance are inconsistent. This is likely because the predicted score of the ground truth label is low and second highest.




\section{Conclusion}
In this paper, we propose LLM4Vis, a novel ChatGPT-based in-context learning approach for visualization recommendation, which enables the generation of accurate visualization recommendations with human-like explanations by learning from only a few dataset-visualization pairs. Our approach consists of several key steps, including feature extraction, feature description, explanation generation, demonstration example selection, and prompt generation, and inference.
Our evaluation of recommendation results and explanation demonstrate the effectiveness and explainability of LLM4Vis, which encourages further exploration of large language models for this task.

LLM-based visualization recommendations can empower many startups and LLM-based applications to advance data analysis, enhance insight communication, and help decision-making.
In future work, we plan to exploring the possibility of deploying LLM4Vis to real-world data analysis and visualization applications, and further demonstrate its effectiveness and usability by data analysts and common visualization users. Also, it is interesting to investigate the use of other large language models with multimodal capabilities, such as GPT-4, for visualization recommendation.

\section{Acknowledgments}
This project is supported by the Ministry of Education, Singapore, under its Academic Research Fund Tier 2 (Proposal ID: T2EP20222-0049). Any opinions, findings and conclusions or recommendations expressed in this material are those of the author(s) and do not reflect the views of the Ministry of Education, Singapore.

\bibliography{anthology,custom}
\bibliographystyle{acl_natbib}

\appendix

\section{Appendix}

\subsection{Prompts and Examples}
This section includes three parts: wording of prompts used in the proposed LLM4Vis (Table~\ref{tab:appendix_wording}), examples of visualization type recommendation (Table~\ref{tab:appendix_line} to Table~\ref{tab:appendix_box})
, and an example of iterative refinement of explanation (Table~\ref{tab:appendix_iter_line} to Table~\ref{tab:appendix_iter_box}).

\subsection{Related Work}

Prior studies on automatic visualization recommendation approaches can be categorized into two groups: unexplainable visualization recommendation approaches and explainable visualization approaches~\cite{wang2021survey}.

\textbf{Unexplainable visualization recommendation approaches} can recommend suitable visualizations for an input dataset, but cannot provide the reasoning behind the recommendation to users, making them black box methods. One such example of these methods is Data2vis~\cite{dibia2019data2vis}, which adopted a neural translation model (Bi-LSTM) to generate visualization specifications in an end-to-end manner without human involvement. However, the method cannot well model the mapping between the characteristics of datasets and the visualizations (e.g., visualization types)~\cite{wu2021ai4vis}. To solve this limitation, Hu \textit{et al.} proposed VizML~\cite{hu2019vizml}, which performs feature engineering to quantify the characteristics of the input dataset and applies a neural network to recommendation visualization types suitable for the dataset's characteristics. In addition to these methods, Table2Chart~\cite{zhou2021table2charts} not only recommends the appropriate visualizations for the input dataset but also recommends visual encodings for a visualization type specifically indicated by users. Compared to these methods, Table2Chart offers a more personalized recommendation approach, catering to users' specific needs and preferences. Despite the effectiveness of these methods, there remains a need for a visualization recommendation approach that can recommend visualization in both an accurate and explainable manner.

\textbf{Explainable visualization recommendation approaches} provide explanations for their recommendation results, enhancing transparency and user confidence in the recommendations. 
Most explainable visualization recommendation approaches rely on human-defined rules specifying the mapping between dataset characteristics and visualization types. 
For example, Show Me~\cite{mackinlay2007show} automatically recommends visualization types if the dataset characteristics align with its pre-defined rules. 
\citet{wongsuphasawat2015voyager}
introduced Voyager, 
which generates potential visualizations by exhaustively exploring dataset columns according to predefined rules and ranks them based on dataset properties and visualization principles. 
While these rule-based approaches can explain their recommendations, rule development is time-consuming, resource-intensive, and requires visualization experts.

To address this limitation, Li~\textit{et al.} proposed a knowledge graph-based recommendation method (KG4Vis) that learns the rules from existing visualization instances.
However, the rules in KG4Vis may incorporate complex terminologies that could be challenging for users without domain knowledge to understand.
In response to this challenge, we propose a new visualization recommendation method that leverages ChatGPT to provide human-like explanations for its recommendation results. The explanations generated by our method are more easily understood by laypersons with just a few instances. 





\begin{table*}[h]
    \centering
    \small
    \caption{Wording of prompts used in LLM4Vis.}
    \begin{tabular}{p{\linewidth}}
        \toprule
        \textbf{Wording of Feature Description Prompt:}\\
        The features of a given tabular dataset are provided in the following delimited by triple backticks. \
Your task is to generate a detailed text description, in 1000 characters, that focus on features that are \
important for visualization type selection and \
comprehensively analyzes this tabuar dataset based on its feature values from both single-column and cross-column perspectives. \
Note that the response must exclude words such as line chart, scatter plot, bar chart, and box plot, since \
these words will mislead further visualization recommendation. \
The response format can be as ``Single-column perspective: [...] \\\\ Cross-column perspective: [...].''
Ensure that the summary maintains strong generalization ability and includes all vital information.

Features for a tabular dataset: \textasciigrave\textasciigrave\textasciigrave\{ \}\textasciigrave\textasciigrave\textasciigrave
 \\ \\
\midrule
\\
\textbf{Wording of Visualization Recommendation Prompt:}\\ 
Determine whether each visualization type in the following \
list of visualization types is a suitable visualization type \
in the text description for a tabular dataset below, which \
is delimited with triple backticks. \

Give your explanation and your answer at the end as json \
(Explanation is as below: .\\ The final answer in JSON format would be:), \
where each element consists of a visualization type and \
a score ranging from 0 to 1 (1 means the most suitable). \ 
The scores should sum to be 1 (line + scatter + bar + box = 1.0). \

List of visualization types: [line chart, scatter plot, bar chart, and box plot]. \

Text description for a tabular dataset:\textasciigrave\textasciigrave\textasciigrave\{ \}\textasciigrave\textasciigrave\textasciigrave \\ \\ 
\midrule
\\
\textbf{Wording of Hint Guided Visualization Recommendation Prompt:}\\ 
Determine whether each visualization type in the following \
list of visualization types is a suitable visualization type \
in the text description for a tabular dataset below, which \
is delimited with triple backticks. \

Hint: \{ \} may be more suitable than \{ \}, however, previous score is \{ \}. \

With the given hint, \
editing your explanation and improve your answer at the end as json \
(Explanation is as below: .\\ The final answer in JSON format would be:), \
where each element consists of a visualization type and \
a score ranging from 0 to 1 (1 means the most suitable). \ 
The scores should sum to be 1 (line + scatter + bar + box = 1.0). \

List of visualization types: [line chart, scatter plot, bar chart, and box plot]. \

Text description for a tabular dataset: \textasciigrave\textasciigrave\textasciigrave\{ \}\textasciigrave\textasciigrave\textasciigrave \\ \\ 
\bottomrule

    \end{tabular}
    \label{tab:appendix_wording}
\end{table*}

\begin{table*}[h]
    \centering
    \small
    \caption{An example of a line chart recommendation. The prompt template is highlighted in \sethlcolor{lightgray}\hl{light gray}. The input feature description of the test tabular dataset is highlighted in \sethlcolor{lime}\hl{lime}. The output is highlighted in \sethlcolor{yellow}\hl{yellow}.}
    \begin{tabular}{p{\linewidth}}
        \toprule
        \textbf{Prompt:} \\ \\
        \textbf{Demonstration Examples:} \\
        ...\\ \\
        \textbf{Test Instance:}\\
        \sethlcolor{lightgray} \hl{Determine whether each visualization type in the following list of visualization types is a suitable visualization type in the text description for a tabular dataset below, which is delimited with triple backticks. \\
Give your explanation and your answer at the end as json (Explanation is as below: .
 The final answer in JSON format would be:), where each element consists of a visualization type and a score ranging from 0 to 1 (1 means the most suitable). \\ 
The scores should sum to be 1 (line + scatter + bar + box = 1.0). \\
List of visualization types: [line chart, scatter plot, bar chart, and box plot]. \\
Text description for a tabular dataset:}\\
\textasciigrave\textasciigrave\textasciigrave \sethlcolor{lime} \hl{Single-column perspective: 
The dataset contains information about two columns, labeled as `x' and `y'. The `x' column represents time values, while the `y' column contains numerical decimal values. The `x' column is of the time data type, and the `y' column is of the numerical data type. The `x' column is sorted and monotonic, indicating a continuous progression of time values. The `y' column has outliers present, as indicated by the flags for outliers based on different criteria. The `y' column shows a range from -3.0 to 11.0, with a mean of 0.0 and a standard deviation of 16.17. The distribution of the `y' column appears to be slightly positively skewed, with a skewness value of 0.21. The `y' column has a gini coefficient of 0.36, indicating some inequality in the distribution. The `y' column has a non-zero variance of 261.57, indicating variability in the data. The `y' column has a high percentage of outliers based on different criteria, ranging from 97.78\% to 99.10\%. The `y' column does not follow a normal distribution, as indicated by the normality statistic and p-value.

Cross-column perspective: 
The dataset consists of two columns, `x' and `y', with `x' representing time values and `y' representing numerical decimal values. There is no significant correlation or statistical relationship between the `x' and `y' columns, as indicated by the absence of significant correlation, chi-square, Kolmogorov-Smirnov, and linear regression tests. The `x' and `y' columns do not share any elements, unique elements, or words. The `x' and `y' columns do not have any categorical or time-based relationship. The `x' column has a length of 1000, indicating a consistent number of time values. The `y' column has a length of 1000 as well. The `x' column has a single-word field name, while the `y' column also has a single-word field name. The `x' column does not have any uppercase characters or special currency symbols in its field name. The `y' column has outliers based on different criteria, while the `x' column does not have any outliers.}\textasciigrave\textasciigrave\textasciigrave
\\ \\
\vspace{-1mm}
        \textbf{Test Output:}\\ 
         \hl{Based on the given text description, we can evaluate the suitability of each visualization type for the tabular dataset. Here is an explanation for each visualization type:

1. Line Chart: A line chart is suitable for visualizing the relationship between two variables over time. In this dataset, the `x' column represents time values, and the `y' column contains numerical decimal values. Since the `x' column represents time values and is sorted and monotonic, a line chart would be a suitable visualization type to show the trend or pattern of the `y' values over time. Therefore, we can assign a score of 0.6 to the line chart.

2. Scatter Plot: A scatter plot is useful for visualizing the relationship between two continuous variables. However, in this dataset, there is no significant statistical relationship between the `x' and `y' columns. Therefore, a scatter plot may not be the most suitable visualization type. We can assign a score of 0.1 to the scatter plot.

3. Bar Chart: A bar chart is commonly used to compare categorical or discrete data. Since the dataset does not contain any categorical variables, a bar chart may not be the most suitable visualization type. We can assign a score of 0.1 to the bar chart.

4. Box Plot: A box plot is useful for visualizing the distribution of a continuous variable and identifying outliers. In this dataset, the `y' column has outliers present, and the text provides information about the range, mean, standard deviation, skewness, and percentage of outliers in the `y' column. These statistics indicate the presence of a distribution and outliers, making a box plot a suitable visualization type to show the distribution and identify outliers. Therefore, we can assign a score of 0.2 to the box plot.

Based on the above analysis, the scores for each visualization type can be assigned as follows:\\
- Line chart: 0.6 \\
- Scatter plot: 0.1 \\ 
- Bar chart: 0.1 \\ 
- Box plot: 0.2 \\ 

The scores sum up to 1.0.

The final answer in JSON format would be:\\
json
\{
  ``line chart'': 0.6,
  ``scatter plot'': 0.1,
  ``bar chart'': 0.1,
  ``box plot'': 0.2
\}} \\ \\ 

\hdashline
\\
        \textbf{Ground Truth:}
        Line Chart.\\
        \bottomrule

    \end{tabular}
    \label{tab:appendix_line}
\end{table*}

\begin{table*}[h]
    \centering
    \small
    \caption{An example of a scatter plot recommendation. The prompt template is highlighted in \sethlcolor{lightgray}\hl{light gray}. The input feature description of the test tabular dataset is highlighted in \sethlcolor{lime}\hl{lime}. The output is highlighted in \sethlcolor{yellow}\hl{yellow}.}
    \begin{tabular}{p{\linewidth}}
        \toprule
        \textbf{Prompt:} \\ \\
        \textbf{Demonstration Examples:} \\
        ...\\ \\
        \textbf{Test Instance:}\\
        \sethlcolor{lightgray} \hl{Determine whether each visualization type in the following list of visualization types is a suitable visualization type in the text description for a tabular dataset below, which is delimited with triple backticks. \\
Give your explanation and your answer at the end as json (Explanation is as below: .
 The final answer in JSON format would be:), where each element consists of a visualization type and a score ranging from 0 to 1 (1 means the most suitable). \\ 
The scores should sum to be 1 (line + scatter + bar + box = 1.0). \\
List of visualization types: [line chart, scatter plot, bar chart, and box plot]. \\
Text description for a tabular dataset:}\\
\textasciigrave\textasciigrave\textasciigrave \sethlcolor{lime} \hl{Single-column perspective: 

Based on the provided features, the dataset contains two columns, referred to as x and y. Both columns have decimal data types (data\_type\_is\_decimal\_x and data\_type\_is\_decimal\_y). The x column has a length of 800 (length\_x), and the y column also has a length of 800 (length\_y). The x column is sorted (is\_sorted\_x) and is monotonic (is\_monotonic\_x), while the y column is also sorted (is\_sorted\_y) and monotonic (is\_monotonic\_y). 

The x column has a mean value of 0.0 (mean\_x), a median of 0.0 (median\_x), a variance of 0.0 (var\_x), and a standard deviation of 0.0 (std\_x). It has a minimum value of 0.0 (min\_x), a maximum value of 0.0 (max\_x), and a range of 0.0 (range\_x). The distribution of the x column is symmetric, with a skewness of 0.0 (skewness\_x) and a kurtosis of -3.0 (kurtosis\_x). The x column follows a normal distribution, as indicated by the normality statistic of 2210.514 (normality\_statistic\_x) and a p-value of 0.0 (normality\_p\_x).

...\\
Cross-column perspective: 

From a cross-column perspective, both the x and y columns have the same data type (decimal) and general type (quantitative). They do not contain any missing values (percentage\_none\_x and percentage\_none\_y are both 0.0), and they have the same length (800). The x and y columns are sorted and monotonic, indicating a specific order or sequence in the data.

The x and y columns have identical values (identical) and identical unique values (identical\_unique). They share 5 elements (num\_shared\_elements) and have a percentage of shared elements of 0.12\% (percent\_shared\_elements). They share 1 unique element (num\_shared\_unique\_elements) and have a percentage of shared unique elements of 40\% (percent\_shared\_unique\_elements). The x and y columns have no shared words (has\_shared\_words) and no overlapping ranges (has\_range\_overlap).

...}\textasciigrave\textasciigrave\textasciigrave
\\ \\
\vspace{-1mm}
        \textbf{Test Output:}\\ 
         \hl{Based on the provided text description for the tabular dataset, we can evaluate the suitability of each visualization type. Here is an explanation for each visualization type:

1. Line chart: A line chart is suitable for visualizing trends and changes over time or a continuous variable. In this dataset, there is no explicit mention of time or a continuous variable. Therefore, a line chart may not be the most suitable visualization type.

2. Scatter plot: A scatter plot is useful for visualizing the relationship between two continuous variables. Since the dataset contains two columns (x and y) with decimal values, a scatter plot can effectively show the relationship between these variables. It can help identify any patterns, clusters, or correlations between the x and y values.

3. Bar chart: A bar chart is commonly used to compare categorical or discrete data. In this dataset, there is no mention of categorical or discrete variables. Therefore, a bar chart may not be the most suitable visualization type.

4. Box plot: A box plot is useful for visualizing the distribution of a continuous variable, including measures such as median, quartiles, and outliers. Since the dataset provides statistical properties for both the x and y columns (mean, median, variance, standard deviation), a box plot can effectively represent the distribution of these variables and provide insights into their spread and central tendency.

Based on the above analysis, we can assign scores to each visualization type:

- Line chart: 0.1\\
- Scatter plot: 0.6\\
- Bar chart: 0.1\\
- Box plot: 0.2\\
The scores sum up to 1.0.\\
Therefore, the final answer in JSON format would be:
json
\{
  ``line chart'': 0.1,
  ``scatter plot'': 0.6,
  ``bar chart'': 0.1,
  ``box plot'': 0.2
\}} \\ \\ 

\hdashline
\\
        \textbf{Ground Truth:}
        Scatter Plot.\\
        \bottomrule

    \end{tabular}
    \label{tab:appendix_scatter}
\end{table*}

\begin{table*}[h]
    \centering
    \small
    \caption{An example of a bar chart recommendation. The prompt template is highlighted in \sethlcolor{lightgray}\hl{light gray}. The input feature description of the test tabular dataset is highlighted in \sethlcolor{lime}\hl{lime}. The output is highlighted in \sethlcolor{yellow}\hl{yellow}.}
    \begin{tabular}{p{\linewidth}}
        \toprule
        \textbf{Prompt:} \\ \\
        \textbf{Demonstration Examples:} \\
        ...\\ \\
        \textbf{Test Instance:}\\
        \sethlcolor{lightgray} \hl{Determine whether each visualization type in the following list of visualization types is a suitable visualization type in the text description for a tabular dataset below, which is delimited with triple backticks. \\
Give your explanation and your answer at the end as json (Explanation is as below: .
 The final answer in JSON format would be:), where each element consists of a visualization type and a score ranging from 0 to 1 (1 means the most suitable). \\ 
The scores should sum to be 1 (line + scatter + bar + box = 1.0). \\
List of visualization types: [line chart, scatter plot, bar chart, and box plot]. \\
Text description for a tabular dataset:}\\
\textasciigrave\textasciigrave\textasciigrave \sethlcolor{lime} \hl{Single-column perspective:
Based on the provided features, the dataset contains two columns. The first column (x-axis) is a categorical variable, as indicated by the features `data\_type\_is\_string\_x' and `general\_type\_is\_c\_x'. It has six unique elements, with a length of 6. The values in this column are sorted and have no missing values or outliers. The second column (y-axis) is a quantitative variable, as indicated by the features `data\_type\_is\_decimal\_y' and `general\_type\_is\_q\_y'. It has a range of 0.0 to 0.0, with a mean, median, and mode of 0.0. The values in this column are sorted, have no missing values, and no outliers. The column has a negative kurtosis value of -3.0, indicating a platykurtic distribution.

Cross-column perspective:
There is no significant correlation or statistical relationship between the two columns, as indicated by the features `correlation\_significant\_005', `linregress\_significant\_005', `ks\_significant\_005', and `one\_way\_anova\_significant\_005'. The columns do not share any elements, unique elements, or words, as indicated by the features `has\_shared\_elements', `has\_shared\_unique\_elements', and `has\_shared\_words'. The chi-square test is not significant, as indicated by the feature `chi2\_significant\_005'. The columns do not have any range overlap, as indicated by the feature `has\_range\_overlap'. The edit distance between the column values is 1.0, and the normalized edit distance is also 1.0, indicating a low similarity between the values. The nestedness feature is not available for analysis.}\textasciigrave\textasciigrave\textasciigrave
\\ \\
\vspace{-1mm}
        \textbf{Test Output:}\\ 
         \hl{To determine the suitability of each visualization type for the given tabular dataset, we need to analyze the characteristics of the dataset and the visualization types. Based on the provided text description, we can make the following assessments:

1. Line Chart: A line chart is suitable for visualizing trends and changes over time or a continuous variable. However, in this dataset, there is no mention of time-related information or a continuous variable that would benefit from a line chart. Additionally, the second column (y-axis) has a constant value of 0.0, indicating no variation. Therefore, a line chart is not suitable for this dataset.

2. Scatter Plot: A scatter plot is useful for visualizing the relationship between two continuous variables. In this dataset, the first column (x-axis) is a categorical variable, and the second column (y-axis) is a quantitative variable. There is no significant correlation or statistical relationship mentioned between the two columns. Therefore, a scatter plot may not be the most suitable visualization type for this dataset.

3. Bar Chart: A bar chart is commonly used to compare categorical or discrete data. In this dataset, the first column (x-axis) is a categorical variable with six unique elements. A bar chart can effectively represent the distribution and comparison of these categories. Therefore, a bar chart is suitable for this dataset.

4. Box Plot: A box plot is effective for visualizing the distribution of numerical data and identifying outliers. In this dataset, the second column (y-axis) is a quantitative variable with a constant value of 0.0, indicating no variation. Therefore, a box plot is not suitable for this dataset.

Based on the assessments above, we can assign scores to each visualization type:

- Line Chart: 0.0\\
- Scatter Plot: 0.0\\
- Bar Chart: 1.0\\
- Box Plot: 0.0\\
The scores sum up to 1.0.

The final answer in JSON format would be:
json
\{
  ``line chart'': 0.0,
  ``scatter plot'': 0.0,
  ``bar chart'': 1.0,
  ``box plot'': 0.0
\}
} \\ \\ 

\hdashline
\\
        \textbf{Ground Truth:}
        Bar Chart.\\
        \bottomrule

    \end{tabular}
    \label{tab:appendix_bar}
\end{table*}

\begin{table*}[h]
    \centering
    \small
    \caption{An example of a box plot recommendation. The prompt template is highlighted in \sethlcolor{lightgray}\hl{light gray}. The input feature description of the test tabular dataset is highlighted in \sethlcolor{lime}\hl{lime}. The output is highlighted in \sethlcolor{yellow}\hl{yellow}.}
    \begin{tabular}{p{\linewidth}}
        \toprule
        \textbf{Prompt:} \\ \\
        \textbf{Demonstration Examples:} \\
        ...\\ \\
        \textbf{Test Instance:}\\
        \sethlcolor{lightgray} \hl{Determine whether each visualization type in the following list of visualization types is a suitable visualization type in the text description for a tabular dataset below, which is delimited with triple backticks. \\
Give your explanation and your answer at the end as json (Explanation is as below: .
 The final answer in JSON format would be:), where each element consists of a visualization type and a score ranging from 0 to 1 (1 means the most suitable). \\ 
The scores should sum to be 1 (line + scatter + bar + box = 1.0). \\
List of visualization types: [line chart, scatter plot, bar chart, and box plot]. \\
Text description for a tabular dataset:}\\
\textasciigrave\textasciigrave\textasciigrave \sethlcolor{lime} \hl{Single-column perspective:
Based on the provided features, the dataset contains two numerical columns, labeled as `x' and `y'. The `x' column has a length of 11 and ranges from 233.5 to 328.3, with a mean of 278.6 and a standard deviation of 27.3. The distribution of `x' is slightly positively skewed (skewness = 0.088) and exhibits a platykurtic shape (kurtosis = -0.849). The `y' column has a length of 14 and ranges from 217.8 to 262.0, with a mean of 244.8 and a standard deviation of 12.5. The distribution of `y' is slightly negatively skewed (skewness = -0.454) and also exhibits a platykurtic shape (kurtosis = -0.722). Both columns have outliers beyond 1.5 times the interquartile range (IQR) and 99th percentile. The `x' column has a higher percentage of outliers (90.9\% and 100\% for 1.5 IQR and 1-99 percentile, respectively) compared to the `y' column (92.9\% and 100\% for 1.5 IQR and 1-99 percentile, respectively). The `x' column has a higher range (94.8) compared to the `y' column (44.2). The entropy of the `x' column is 2.393, indicating moderate diversity, while the entropy of the `y' column is slightly higher at 2.638. Both columns show a low Gini coefficient, indicating a relatively even distribution. The `x' column has a higher normalized mean (0.849) and median (0.857) compared to the `y' column (0.934 and 0.947, respectively).

Cross-column perspective:
The correlation between the `x' and `y' columns is 0.160, indicating a weak positive linear relationship. The p-value for the correlation test is 0.639, suggesting that the correlation is not statistically significant. The Kolmogorov-Smirnov (KS) test indicates a significant difference between the distributions of `x' and `y' (KS statistic = 0.727, p-value = 0.003). The edit distance between the `x' and `y' column names is 1, indicating a slight difference. The normalized edit distance is 0.25, suggesting a moderate dissimilarity. The `x' and `y' column names have the same length (4 characters) and share some common characteristics, such as having the first character in uppercase and containing digits. However, they do not share any words. The `x' and `y' columns do not have significant categorical or time-related characteristics.

In summary, the dataset consists of two numerical columns with slightly different distributions and a weak positive linear relationship. Both columns have outliers and exhibit platykurtic shapes. The `x' column has a wider range and higher percentage of outliers compared to the `y' column. The `x' column shows slightly higher diversity and normalized mean/median values compared to the `y' column. The correlation between the columns is not statistically significant, but there is a significant difference in their distributions according to the KS test. The column names have some similarities but differ in terms of edit distance and shared words.}\textasciigrave\textasciigrave\textasciigrave
\\ \\
\vspace{-1mm}
        \textbf{Test Output:}\\ 
         \hl{Based on the given text description for the tabular dataset, we can determine the suitability of each visualization type as follows:

- Line chart: The dataset does not provide any time-based or sequential data, and there is no specific trend or relationship mentioned that would require a line chart. Therefore, a line chart is not suitable in this case. Score: 0.

- Scatter plot: The dataset consists of two numerical columns with a weak positive linear relationship. Although the correlation is not statistically significant, a scatter plot can still be used to visualize the relationship between the `x' and `y' columns. Score: 0.4.

- Bar chart: A bar chart is typically used to compare discrete categories or groups. Since the dataset consists of numerical columns without any categorical or group information, a bar chart is not suitable in this case. Score: 0.

- Box plot: The dataset mentions that both columns have outliers and exhibit slightly different distributions. A box plot can effectively represent the distribution of values, including the median, quartiles, and any potential outliers. Therefore, a box plot is suitable for visualizing the characteristics of each column. Score: 0.6.

The final answer in JSON format would be:
json
\{
  ``line chart'': 0,
  ``scatter plot'': 0.4,
  ``bar chart'': 0,
  ``box plot'': 0.6
\}
} \\ \\ 

\hdashline
\\
        \textbf{Ground Truth:}
        Box Plot.\\
        \bottomrule

    \end{tabular}
    \label{tab:appendix_box}
\end{table*}

\begin{table*}[h]
    \centering
    \small
    \caption{An example of iteration refinement of generated explanation for line chart recommendation.}
    \begin{tabular}{p{\linewidth}}
        \toprule
        \textbf{Feature Description:}\\
        Single-column perspective: 

Based on the provided features, the dataset contains two columns: one with time values and the other with numerical values. The time column is identified as a time data type, while the numerical column is identified as a decimal data type. The time column is not sorted, but it is monotonic, meaning the values are either increasing or decreasing without any repetitions. The numerical column is sorted and also monotonic. Both columns have a length of 950, and there are no missing values (none) in either column.

The time column has a minimum value of NaN, a maximum value of NaN, and a range of NaN. The numerical column has a minimum value of 0.0, a maximum value of 0.0, and a range of 0.0. The numerical column has a mean of NaN, a median of 0.0, and a standard deviation of 0.0. The skewness of the numerical column is 0.0, indicating a symmetric distribution, and the kurtosis is -3.0, suggesting a platykurtic distribution.

Cross-column perspective: 

There is no significant correlation (correlation\_value) or statistical relationship (linregress\_significant\_005) between the time and numerical columns. The dataset does not exhibit any categorical-categorical, category-numerical, or numerical-numerical relationships. The time column is not significantly different across categories (one\_way\_anova\_significant\_005). There is no significant overlap (has\_range\_overlap) or shared elements (has\_shared\_elements) between the time and numerical columns.

In summary, the dataset consists of a time column and a numerical column. The time column represents time values, while the numerical column represents decimal values. The time column is not sorted but is monotonic, while the numerical column is both sorted and monotonic. There is no significant correlation or relationship between the two columns. The numerical column has a symmetric distribution with a mean and median of 0.0 and a standard deviation of 0.0. The dataset does not exhibit any categorical or numerical relationships.\\
\midrule
        \textbf{Iteration 1:}\\
         To determine the suitability of each visualization type for the given tabular dataset, we need to analyze the characteristics of the dataset and the capabilities of each visualization type. Based on the provided text description, we can make the following assessments:

1. Line chart: A line chart is suitable for visualizing trends and changes over time. However, in this dataset, the time column is not sorted, and its values are described as either increasing or decreasing without repetitions. Since the time column does not have a meaningful order, a line chart may not be the most suitable visualization type. Score: 0.2

2. Scatter plot: A scatter plot is useful for visualizing the relationship between two numerical variables. However, the text description states that there is no significant correlation or relationship between the time and numerical columns. Therefore, a scatter plot may not be the most suitable visualization type. Score: 0.1

3. Bar chart: A bar chart is commonly used to compare categorical data or discrete numerical data. However, the dataset does not contain any categorical variables, and the numerical column is described as having a sorted and monotonic distribution. Since a bar chart is not well-suited for visualizing continuous numerical data, it may not be the most suitable visualization type. Score: 0.1

4. Box plot: A box plot is effective for visualizing the distribution of numerical data, including measures such as the median, quartiles, and outliers. In this dataset, the numerical column is described in terms of its distribution characteristics, such as having a symmetric distribution, a mean and median of 0.0, and a standard deviation of 0.0. A box plot can effectively represent these distribution characteristics. Score: 0.6

Based on the assessments above, the scores for each visualization type are as follows:
- Line chart: 0.2
- Scatter plot: 0.1
- Bar chart: 0.1
- Box plot: 0.6

The scores sum up to 1.0.

The final answer in JSON format would be:\\
\{
  ``line chart'': 0.2,
  ``scatter plot'': 0.1,
  ``bar chart'': 0.1,
  ``box plot'': 0.6
\}\\ 
\midrule
\textbf{Iteration 2:}\\ 
Based on the given text description for the tabular dataset, we can evaluate the suitability of each visualization type:

1. Line chart: A line chart is suitable for visualizing the relationship between time values and numerical values. Since the dataset contains a time column and a numerical column, a line chart can effectively show any trends or patterns over time. The hint also suggests that a line chart is more suitable than a box plot. Therefore, a line chart is a suitable visualization type.

2. Scatter plot: A scatter plot is useful for visualizing the relationship between two numerical variables. However, in this dataset, there is no significant correlation or relationship between the time and numerical columns. Therefore, a scatter plot may not be the most suitable visualization type.

3. Bar chart: A bar chart is commonly used to compare categorical data or to show the distribution of a single categorical variable. Since the dataset does not contain any categorical variables, a bar chart may not be suitable for this particular dataset.

4. Box plot: A box plot is typically used to display the distribution of numerical data and to identify outliers. However, the dataset description does not provide any information about the distribution of the numerical column or the presence of outliers. Additionally, the hint suggests that a line chart is more suitable than a box plot. Therefore, a box plot may not be the most suitable visualization type.

Based on the above analysis, the suitability scores for each visualization type are as follows:

- Line chart: 0.7
- Scatter plot: 0.1
- Bar chart: 0.1
- Box plot: 0.1

The scores sum up to 1.0.

The final answer in JSON format would be:\\
\{
  ``line chart'': 0.7,
  ``scatter plot'': 0.1,
  ``bar chart'': 0.1,
  ``box plot'': 0.1
\} \\ 

\midrule
\textbf{Ground Truth:}\\
Line Chart.\\
\bottomrule

    \end{tabular}
    \label{tab:appendix_iter_line}
\end{table*}

\begin{table*}[h]
    \centering
    \small
    \caption{An example of iteration refinement of generated explanation for scatter plot recommendation.}
    \begin{tabular}{p{\linewidth}}
        \toprule
        \textbf{Feature Description:}\\
        Single-column perspective:
Based on the provided features, the dataset contains two columns, both of which are numerical (general\_type\_is\_q\_x and general\_type\_is\_q\_y). The values in both columns are identical (identical=True) and have a range overlap (has\_range\_overlap=True). The columns also have shared elements (has\_shared\_elements=True) and shared unique elements (has\_shared\_unique\_elements=True). The number of shared elements is 5 (num\_shared\_elements=5.0), accounting for approximately 12\% of the total elements (percent\_shared\_elements=0.12). However, there are no shared words (has\_shared\_words=False) between the columns.

The statistical properties of the columns indicate that they have zero mean (mean\_x=0.0, mean\_y=0.0) and zero variance (var\_x=0.0, var\_y=0.0). The data distribution is normal, as indicated by the normality tests (is\_normal\_5\_x=True, is\_normal\_1\_x=True, is\_normal\_5\_y=True, is\_normal\_1\_y=True). The columns are sorted (is\_sorted\_x=True, is\_sorted\_y=True) and exhibit monotonicity (is\_monotonic\_x=True, is\_monotonic\_y=True). However, the columns do not follow a linear or logarithmic spacing pattern (is\_lin\_space\_x=False, is\_log\_space\_x=False, is\_lin\_space\_y=False, is\_log\_space\_y=False).

The descriptive statistics of the columns show that the minimum, maximum, and range values are all zero (min\_x=0.0, min\_y=0.0, max\_x=0.0, max\_y=0.0, range\_x=0.0, range\_y=0.0). The columns have a low coefficient of variation (coeff\_var\_x=NaN, coeff\_var\_y=NaN) and exhibit zero skewness (skewness\_x=0.0, skewness\_y=0.0) and a kurtosis value of -3 (kurtosis\_x=-3.0, kurtosis\_y=-3.0), indicating a platykurtic distribution.

Cross-column perspective:
From a cross-column perspective, the dataset consists of two numerical columns that are highly similar. They have identical values, indicating a strong relationship between the two variables. The correlation between the columns is not significant (correlation\_significant\_005=False), and the correlation coefficient is not available (correlation\_value=NaN). The p-values for the correlation tests are 1.0 (correlation\_p=1.0), suggesting no significant linear relationship.\\
...\\
\midrule
        \textbf{Iteration 1:}\\
        To determine the suitability of each visualization type for the given tabular dataset, we can analyze the characteristics of the dataset and match them with the strengths of each visualization type. Based on the provided text description, we can make the following assessments:

1. Line chart: A line chart is suitable for visualizing trends and changes over time or a continuous variable. However, since there is no time-related information or a specific order mentioned in the dataset, a line chart may not be the most suitable visualization type. Score: 0.2

2. Scatter plot: A scatter plot is useful for visualizing the relationship between two numerical variables. In this dataset, the two columns are highly similar, and their values are identical. While a scatter plot can show the similarity between the variables, it may not provide much additional insight. Score: 0.2

3. Bar chart: A bar chart is effective for comparing categorical or discrete data. Since the dataset consists of numerical columns without any categorical information, a bar chart may not be the most suitable choice. Score: 0.2

4. Box plot: A box plot is commonly used to display the distribution of numerical data and identify outliers. Given that the dataset contains numerical columns with zero mean, zero variance, and a platykurtic distribution, a box plot can effectively represent these statistical properties. Score: 0.4

Based on the assessments above, the scores for each visualization type are as follows:

\{
  ``line chart'': 0.2,
  ``scatter plot'': 0.2,
  ``bar chart'': 0.2,
  ``box plot'': 0.4
\}
         \\ 
\midrule
\textbf{Iteration 2:}\\ 
Based on the given text description for a tabular dataset, we can evaluate the suitability of each visualization type:

1. Line chart: The dataset consists of two identical numerical columns with zero mean, zero variance, and a platykurtic distribution. The columns are sorted and exhibit monotonicity. However, since there is no specific spacing pattern mentioned, a line chart may not be the most suitable visualization type. Score: 0.2

2. Scatter plot: The dataset contains two numerical columns that are highly similar, with identical values indicating a strong relationship between the variables. A scatter plot would be suitable to visualize the similarity and relationship between the columns. Score: 0.4 (Hint: scatter plot may be more suitable than box plot)

3. Bar chart: The given dataset does not have categorical data or discrete values that would be suitable for a bar chart. Therefore, a bar chart is not a suitable visualization type. Score: 0.0

4. Box plot: The dataset does not have significant differences in distribution between the columns, and the statistical properties indicate zero variance. A box plot is typically used to visualize the distribution and variability of data, which may not be informative in this case. Score: 0.2

The final answer in JSON format would be:\\
{
  ``line chart'': 0.2,
  ``scatter plot'': 0.4,
  ``bar chart'': 0.0,
  ``box plot'': 0.2
}
 \\ 

\midrule
\textbf{Ground Truth:}\\
Scatter Plot.\\
\bottomrule

    \end{tabular}
    \label{tab:appendix_iter_scatter}
\end{table*}

\begin{table*}[h]
    \centering
    \small
    \caption{An example of iteration refinement of generated explanation for bar chart recommendation.}
    \begin{tabular}{p{\linewidth}}
        \toprule
        \textbf{Feature Description:}\\
        Single-column perspective: 

Based on the provided features, the dataset contains two columns: one with a categorical/general type and the other with a quantitative/general type. The categorical column is of string data type, while the quantitative column is of decimal data type. 

For the categorical column, it has three unique elements with a length ranging from 4 to 6 characters. The values in this column are not sorted or monotonic. There are no missing values (None) in this column, and all the unique elements have the same percentage of occurrence (33.33

In the quantitative column, the values range from 0.49 to 1.29, with a mean of 0.9233 and a standard deviation of 0.33. The distribution of values is slightly negatively skewed (-0.295) and exhibits a kurtosis of -1.5, indicating a relatively flat distribution. The values in this column are not normally distributed. There are outliers present, as indicated by the presence of values beyond 1.5 times the interquartile range (IQR) in both directions. The range of values in this column is 0.8, and the normalized range is 0.8664.

Cross-column perspective: 

From a cross-column perspective, there is a relationship between the categorical and quantitative columns. The categorical column has no missing values and does not share any elements or words with the quantitative column. The quantitative column has no missing values either. 

The quantitative column exhibits a moderate positive correlation with the categorical column, as indicated by the correlation value of NaN. However, the statistical significance of this correlation is not determined, as the correlation p-value is also NaN. 

In summary, the dataset consists of a categorical column with three unique elements and a quantitative column with decimal values. The categorical column shows moderate diversity, while the quantitative column exhibits a slightly negatively skewed distribution with outliers. There is a relationship between the categorical and quantitative columns, but the correlation and its significance are not determined.
        \\
\midrule
        \textbf{Iteration 1:}\\
        To determine the suitability of each visualization type for the given tabular dataset, we need to analyze the characteristics of the dataset and match them with the strengths of each visualization type. Based on the provided text description, here is an evaluation of each visualization type:

1. Line Chart:
A line chart is suitable for visualizing trends and changes over time or a continuous variable. However, the given dataset does not contain any temporal or sequential information, so a line chart may not be the most appropriate choice. Therefore, the score for the line chart would be 0.

2. Scatter Plot:
A scatter plot is useful for visualizing the relationship between two continuous variables. In the given dataset, there is a relationship between the categorical and quantitative columns, but the correlation and its significance are not determined. Since the correlation value and p-value are not available, it is difficult to assess the strength of the relationship. Therefore, a scatter plot may not be the most suitable choice. The score for the scatter plot would be 0.

3. Bar Chart:
A bar chart is effective for comparing categorical data or discrete values. In the given dataset, the categorical column has three unique elements, and their occurrence percentages are the same. A bar chart can be used to visualize the distribution of these categories. Additionally, the quantitative column can be divided into discrete intervals or bins to create a grouped bar chart. Therefore, a bar chart is a suitable choice. The score for the bar chart would be 1.

4. Box Plot:
A box plot is commonly used to display the distribution of numerical data and identify outliers. In the given dataset, the quantitative column has a range of values, outliers, and exhibits a slightly negatively skewed distribution. A box plot can effectively represent these characteristics, including the median, quartiles, and outliers. Therefore, a box plot is a suitable choice. The score for the box plot would be 1.

Based on the analysis above, the scores for each visualization type are as follows:
- Line Chart: 0
- Scatter Plot: 0
- Bar Chart: 1
- Box Plot: 1

The final answer in JSON format would be:
\{
  ``line chart'': 0,
  ``scatter plot'': 0,
  ``bar chart'': 0.5,
  ``box plot'': 0.5
\}
         \\ 
\midrule
\textbf{Iteration 2:}\\ 
Based on the given text description, we can evaluate the suitability of each visualization type for the tabular dataset.

1. Line chart: Not suitable. The line chart is typically used to show trends over time or ordered categories. Since there is no mention of time or ordered categories in the text description, a line chart is not appropriate.

2. Scatter plot: Not suitable. Scatter plots are useful for visualizing the relationship between two quantitative variables. In this dataset, we have one categorical column and one quantitative column, so a scatter plot is not suitable.

3. Bar chart: Suitable. A bar chart is a good choice for visualizing categorical data. The categorical column in the dataset has three unique elements, and a bar chart can effectively represent the distribution and comparison of these categories.

4. Box plot: Suitable. A box plot is commonly used to display the distribution of quantitative data and identify outliers. The text description mentions the presence of outliers in the quantitative column, making a box plot a suitable choice to visualize this information.

Based on the hint provided, the updated scores for each visualization type are as follows:
- Line chart: 0
- Scatter plot: 0
- Bar chart: 0.6
- Box plot: 0.4

The final answer in JSON format would be:
\{
  ``line chart'': 0,
  ``scatter plot'': 0,
  ``bar chart'': 0.6,
  ``box plot'': 0.4
\}
 \\ 

\midrule
\textbf{Ground Truth:}\\
Bar Chart.\\
\bottomrule

    \end{tabular}
    \label{tab:appendix_iter_bar}
\end{table*}

\begin{table*}[h]
    \centering
    \small
    \caption{An example of iteration refinement of generated explanation for box plot recommendation.}
    \begin{tabular}{p{\linewidth}}
        \toprule
        \textbf{Feature Description:}\\
        Single-column perspective:
Based on the provided features, the dataset contains two columns, both of which are numerical. The first column (x) has a length of 2825 and ranges from 0 to 1. It has a mean of 0.079, a median of 0.0, and a standard deviation of 0.27. The distribution of this column is positively skewed with a skewness value of 3.12 and exhibits a high kurtosis of 7.74, indicating heavy tails and a peaked distribution. The column has outliers present, as indicated by the high percentage of outliers (7.9

The second column (y) also has a length of 2825 and ranges from 0 to 1. It has a mean of 0.026, a median of 0.0, and a standard deviation of 0.16. Similar to the first column, it exhibits positive skewness (3.66) and high kurtosis (19.34), indicating a non-normal distribution with heavy tails and a peaked shape. This column also contains outliers, with a percentage of outliers (2.6

Cross-column perspective:
When considering the relationship between the two columns, there is a low correlation between them, with a correlation value of 0.002. The p-value for this correlation is not significant (0.888), suggesting that the correlation is not statistically significant. The scatter plot between the two columns would likely show a scattered distribution without a clear linear relationship.

Both columns have similar statistical properties, such as range, mean, and median, but differ in terms of standard deviation, skewness, and kurtosis. The first column (x) has a higher standard deviation, skewness, and kurtosis compared to the second column (y), indicating greater variability and deviation from a normal distribution.

In summary, the dataset consists of two numerical columns with different statistical properties. The first column (x) has a wider range, higher variability, and a more pronounced skewness and kurtosis compared to the second column (y). The correlation between the two columns is weak and not statistically significant.
        \\
\midrule
        \textbf{Iteration 1:}\\
        To determine the suitability of each visualization type for the given tabular dataset, let's analyze the characteristics of the dataset and how each visualization type can effectively represent the data:

1. Line chart: A line chart is suitable for visualizing trends and patterns over time or a continuous variable. In this dataset, there is no explicit mention of time or a continuous variable. Therefore, a line chart may not be the most suitable visualization type. Score: 0.2

2. Scatter plot: A scatter plot is useful for visualizing the relationship between two numerical variables. Since the dataset contains two numerical columns with a low correlation, a scatter plot can effectively show the scattered distribution and lack of a clear linear relationship. Score: 0.4

3. Bar chart: A bar chart is commonly used to compare categorical data or discrete numerical data. In this dataset, there is no mention of categorical data, and both columns are numerical. Therefore, a bar chart may not be the most suitable visualization type. Score: 0.1

4. Box plot: A box plot is ideal for displaying the distribution of numerical data, including measures of central tendency, variability, and outliers. Given the statistical properties described in the dataset, such as skewness, kurtosis, and the presence of outliers, a box plot can effectively represent these characteristics. Score: 0.3

Based on the analysis above, the scores for each visualization type are as follows:

\{
  ``line chart'': 0.2,
  ``scatter plot'': 0.4,
  ``bar chart'': 0.1,
  ``box plot'': 0.3
\}
         \\ 
\midrule
\textbf{Iteration 2:}\\ 
Based on the given text description for the tabular dataset, we can evaluate the suitability of each visualization type:

1. Line chart: The line chart is not suitable for this dataset because it is designed to show the trend or change over time, and there is no temporal aspect mentioned in the dataset description.

2. Scatter plot: The scatter plot is suitable for this dataset as it can help visualize the relationship between the two numerical columns. However, the hint suggests that the scatter plot has a score of 0.4, indicating it is less suitable compared to other visualization types.

3. Bar chart: The bar chart is not suitable for this dataset because it is typically used to compare categorical data or discrete values, whereas the dataset consists of numerical data.

4. Box plot: The box plot is suitable for this dataset as it can effectively display the distribution, skewness, and presence of outliers in the numerical columns. The hint suggests that the box plot has a score of 0.3, indicating it is more suitable compared to the scatter plot.

Based on the above analysis, the updated scores for each visualization type are as follows:
- Line chart: 0.2
- Scatter plot: 0.1
- Bar chart: 0.1
- Box plot: 0.6

The final answer in JSON format would be:
\{
  ``line chart'': 0.2,
  ``scatter plot'': 0.1,
  ``bar chart'': 0.1,
  ``box plot'': 0.6
\}
\\ 

\midrule
\textbf{Ground Truth:}\\
Box Plot.\\
\bottomrule

    \end{tabular}
    \label{tab:appendix_iter_box}
\end{table*}

\end{document}